\documentclass[pre,twocolumn,showpacs]{revtex4}

\usepackage{graphicx}
                                                                                
\begin{document}
\title[Version 1]{Boundary effects on localized structures in spatially
extended systems}
                                                                                
\author{A. Yadav}
 \email{yadav@phys.lsu.edu}
\author{D. A. Browne}
\affiliation{Department of Physics and Astronomy,
Louisiana State University,
Baton Rouge, LA 70803--4001}
                                                                                
\pacs{82.40.-g,05.70.Ln}
\begin{abstract}
We present a general method of analyzing the influence of finite size and 
boundary effects on the dynamics of localized solutions of non-linear
spatially extended systems. The dynamics of localized structures in infinite
systems involve solvability conditions that require projection onto a 
Goldstone mode. Our method works by extending the solvability conditions to
finite sized systems, by incorporating the finite sized modifications of the 
Goldstone mode and associated nonzero eigenvalue. We apply this method to the 
special case of non-equilibrium domain walls under the influence of Dirichlet 
boundary conditions in a parametrically forced complex Ginzburg Landau 
equation, where we examine exotic nonuniform domain wall motion due to the 
influence of boundary conditions.
\end{abstract}
\maketitle
                                                                                
\section{Introduction}

Models of non-linear spatially extended systems exhibit a variety of spatial
and temporal pattern forming phenomena. A subclass of these patterns are
spatially localized structures \cite{rev1} that include pulses, solitons,
fronts, and domain walls. The standard analysis of these localized structures
assumes that, on large length and time scales, they can be 
treated as ``coherent objects'' \cite{rev1}, with effective parameters like 
position, and velocity attributed to them. A perturbative expansion about 
this isolated coherent object profile is then used to understand 
its response to external forces, interaction with 
other localized structures \cite{ephlick1,ephlick}, noise, or internal 
instabilities \cite{meron4,skyrabin}. Perturbative calculations reduce the 
original non-linear problem to a series of linear problems that require 
consistency criteria known as solvability conditions for their solution. 
Typically, the solution of a linear equation $L \phi= \psi$, requires the 
orthogonality of $\psi$ to the zero modes $\chi$, ie., $(\psi,\chi)=0$, in the 
null space of the adjoint homogeneous problem $L^{\dag} \chi=0$.

Often, the symmetries in a particular system are responsible for the
zero modes of the operators obtained after a perturbative expansion. 
For instance, since a localized structure profile and the same 
profile translated infinitesimally are both solutions of the underlying 
non-linear equation, the difference of the two profiles provides a zero 
(neutral or Goldstone) mode. Strictly, the zero mode is the derivative of the 
localized structure profile, and the underlying symmetry is translation
invariance. Zero modes extracted from symmetry arguments may then be employed
straightforwardly into solvability integrals.

The argument above, based on translational invariance, works if the system 
size is infinite. For a localized 
structure near a system boundary, due to the relevant 
boundary conditions that have to be imposed there, the localized structure 
solution and its infinitesimally translated counterpart are no longer 
solutions of the same equation. Hence, translational invariance is broken. 
Therefore, in this case, one has to contend, not only
with the incorporation of the boundary data into the solvability condition, but
also the appropriate treatment of broken translation invariance.

Most treatments of localized structures follow analytical techniques that
fall in the realm of moving boundary approximations \cite{fife}. A common
feature to these approximations, for instance, in excitable waves 
\cite{meronrp}, or bistable fronts \cite{meron5, siam1, meron2}, is the 
separation of the description of the localized structure into an 
``inner region'' and ``outer region''. The inner region, characterized 
by short spatial scales
and fast time scales, captures the internal dynamics of the 
localized structure. In contrast, the dynamics of the localized structure 
as a whole is captured by
the long spatial and time scales comprising the outer problem. The solvability
integrals in moving boundary type approximations occur in the inner 
problem. Since it is the fields in the outer region that mediate the 
interaction with the boundary \cite{meron1,yadav}, the boundary data is not 
incorporated into solvability conditions arising in the inner problem. 
There are ample situations however, where it may not be possible to have 
separate inner and outer regions of a localized structure by manipulating 
relevant system parameters \cite{Coullet}. In such cases, the boundary 
data must be directly incorporated into the solvability condition.

In this paper, through an appropriately chosen adjoint operator 
$L^{\dag}$ defined for the semi-infinite system (localized structure near 
a boundary), we develop techniques that not only include the boundary data
into the solvability condition, but also directly incorporate the effects
of broken translational invariance into it. We accomplish this by extending
the definition of the Goldstone mode to include the possibility that the
corresponding eigenvalue be non-zero, with its magnitude dependent on how
close the localized structure is to the boundary. This leads further to a 
modified solvability criteria.

As a case study, we develop our techniques in the context of reaction-diffusion
systems and apply it to non-equilibrium domain walls (fronts) found in
bistable regimes. In bistable reaction-diffusion systems, fronts connecting
the two homogeneous steady states can undergo a bifurcation, called a front
bifurcation, where a stationary Ising front loses stability to two 
counter-propagating Bloch fronts\cite{Coullet}. This bifurcation can be 
regarded as an internal instability of the Ising front, the localized 
structure about which a perturbative expansion is carried out to obtain the
propagating Bloch wall solution. This bifurcation, also known as the 
Ising-Bloch bifurcation, has been observed in several systems, like chemical 
reactions \cite{meron4, haas, Li} and also in liquid crystals \cite{Frisch,kai}. 

In a recent work \cite{yadav}, we examined the influence of boundaries on
Ising-Bloch fronts in a FitzHugh-Nagumo (FHN) reaction diffusion model. We 
were able to derive order parameter equations (OPE) for front dynamics, where 
the fronts were perturbed by the imposition of Dirichlet and possibly other 
boundary conditions at the boundaries. This derivation for the two component 
FHN model required restrictive assumptions about the relative size of the 
fronts for the two concentration fields, allowing for the use of moving boundary
approximation like singular perturbation
methods detailed in \cite{siam1,meron2}. These singular perturbation
techniques are quite versatile, predicting exotic phenomena like front 
reversal, trapping, and oscillation at the boundary. However, as observed 
earlier, we wish to examine the effects of boundary data on localized 
structures, where moving boundary type approximations are not applicable, 
and the explicit incorporation of boundary data in a solvability condition is 
required.

In the next section we discuss the extension of the solvability condition to
incorporate boundary data and broken translational invariance
via the extension of the Goldstone mode in a generic system exhibiting a 
localized structure. In Sec. III, we describe the modification of the
slow manifold of a generic Ising-Bloch front due to boundary effects.
In Sec. IV, we 
apply our method of solvability condition extension to study the effects of 
finite size and Dirichlet boundary conditions on the dynamics of Ising-Bloch 
fronts in a parametrically forced complex Ginzburg Landau equation (CGLE)
\cite{Coullet,skyrabin}. An important reason behind this choice is its 
experimental context, modeling Ising-Bloch fronts in Liquid crystals subjected 
to rotating magnetic fields \cite{Frisch,kai}. Liquid crystal systems are
ideal candidates to study boundary effects, as lateral boundary conditions 
may be imposed in a controlled manner by appropriate electric fields 
\cite{book}. Another experimental test bed is presented Ref.~\cite{bode2}, 
in the form of coupled non-linear electrical oscillators, where the application
of boundary conditions requires a minor and straightforward variation of the
original circuit. In Sec. V we discuss in detail the implications of the 
derived order parameter equations for the parametrically forced CGLE. In 
Sec. VI we present our conclusions.

\section{Goldstone modes and solvability criteria}
Consider a general non-linear PDE,
\begin{eqnarray}
\partial_t U= {\cal L} U +N(U),
\label{eq:gone}
\end{eqnarray}
where $U(x,t)$ is the solution vector, ${\cal L}$ are the linear terms, 
and $N(U)$ are the non-linear terms. Let $U_0(x)$ be a stationary 
localized solution of Eq.~(\ref{eq:gone}), with the asymptotic behavior
$U(x)\rightarrow 0; x\rightarrow \pm \infty$. In principle, $U_0(x)$ also
encompasses uniformly translating localized structures, which are stationary 
in a co-moving frame.

Due to translational invariance in the system, one has $A(x)=U_{0x}$, the
derivative with respect to $x$ of the localized structure profile, as the 
zero eigenvalue (neutral or Goldstone) mode of the operator 
$\pounds={\cal L}+N^{\prime}(U_0)$. Also, it is reasonable to expect that
due to translational invariance $\pounds^{\dag}$ has a 
corresponding zero eigenvector, given by the solution of 
$\pounds^{\dag} A^{\dag}=0$. A detailed discussion of this issue may be found 
in \cite{sarlos} and the references therein.

While examining the stability of $A=U_{0x}$ to perturbations, which may 
include a small external perturbation $p(U,x)$ added onto Eq.~(\ref{eq:gone}), 
one obtains,
\begin{eqnarray}
& &[{\cal L}+N^{\prime}(U_0)]\delta U =f;~~~~~~\nonumber\\
f&=&\partial_t (\delta U)-\{N^{\prime \prime}(U_0){(\delta U)}^2/2+p(U_0,x)
\nonumber\\&+&p^{\prime}(U_0,x)\delta U 
+p^{\prime \prime}(U_0,x){(\delta U)}^2/2 +{\cal O}[{(\delta U)}^3]\},
\label{eq:gfive}
\end{eqnarray}
where $\delta U$ is the small deviation from the localized structure profile.
Realizing that the operator $\pounds={\cal L}+N^{\prime}(U_0)$ has a 
Goldstone mode, the solvability of Eq.~(\ref{eq:gfive}) requires, 
\begin{eqnarray}
(f,A^{\dag})=0.
\label{eq:gsix}
\end{eqnarray}
The brackets indicate an inner product or the projection of the dynamical
terms $f$ onto the Goldstone mode (its corresponding adjoint) $A^{\dag}$. 
Equation.~(\ref{eq:gsix}) represents the generic response of a localized 
structure to a wide variety of perturbations, both internal and external.
From an informal and intuitively appealing point of view, the Goldstone mode
with its associated zero eigenvalue is a slow (relevant) mode, which coupled 
with other slow modes in the system, should dominate the dynamics. The 
projection in Eq.~(\ref{eq:gsix}) is a formal prescription to capture 
this slow dynamics.

Let a localized structure be located near a boundary at $x=-l$, with the
origin fixed at the position of the localized structure. 
Although, $A^{\dag}$ is still a solution of $\pounds^{\dag} A^{\dag}=0$ 
in this case, it does not
assume the homogeneous boundary value $A^{\dag}(-l)=0$. 
Consequently, $A^{\dag}$ is no longer the zero eigenvector of the adjoint 
homogeneous problem in the semi-finite interval $[-l,\infty]$. 
However, we still expect $A^{\dag}$ to play a central role in the dynamics of 
the localized structure, all be it in a slightly modified form 
$A^{\dag}_l=A^{\dag}+\delta A^{\dag}_l$. The subscript $l$ denotes the 
proximity of 
the localized structure to the boundary, and $\delta A^{\dag}_l$ is a 
proximity dependent correction to $A^{\dag}$. We require that in the 
limit $l\rightarrow \infty$, $A^{\dag}_l\rightarrow A^{\dag}$, and 
$\delta A^{\dag}_l\rightarrow 0$. 
This requirement is reasonable on physical grounds.
The slow dynamics of the localized structure far away from the boundary 
involves $A^{\dag}$ as a relevant constituent by virtue of it being a slow 
mode. As the localized structure gradually nears the boundary, we still expect 
$A^{\dag}$, in its modified form $A^{\dag}_l$, to be the relevant (slow) 
constituent of the dynamics. 

$A^{\dag}_l$ may be determined in two possible ways. Firstly, we may extract 
$A^{\dag}_l$ as the solution of
\begin{eqnarray} 
\pounds^{\dag} A^{\dag}_l=0,~~A^{\dag}_l(-l)=0,~~A^{\dag}_l(\infty)=0,
\label{eq:gseven}
\end{eqnarray}
with the implication that $A^{\dag}_l=A^{\dag}+\delta A^{\dag}_l$ is still a 
zero eigenvector in the finite system. Or we may 
extract $A^{\dag}_l$ as a solution of
\begin{eqnarray}
\pounds^{\dag} A^{\dag}_l=\lambda_l A^{\dag}_l,~~A^{\dag}_l(-l)=0.~~
A^{\dag}_l(\infty)=0.
\label{eq:geight}
\end{eqnarray}  
Thus, as the localized structure gradually closes in on a 
boundary, the zero eigenvector $A^{\dag}$ is modified to $A^{\dag}_l$, and 
the zero eigenvalue gradually migrates away from zero, assuming the 
value $\lambda_l$. Hence, as $l \rightarrow \infty$, 
$\lambda_l \rightarrow 0$, and $A^{\dag}_l \rightarrow A$.

The first scenario is easily discarded using uniqueness arguments. If
Eq.~(\ref{eq:gseven}) is obeyed, then $\delta A^{\dag}_l$ 
should obey, $\pounds^{\dag} \delta A^{\dag}_l = 0, 
\delta A^{\dag}_l(-l)=-A^{\dag}(-l), \delta A^{\dag}_l(\infty)=0$, with the
unique solution $\delta A^{\dag}_l=-A^{\dag}$. Therefore, since 
$A^{\dag}_l=A^{\dag}+\delta A^{\dag}_l$, 
Eq.~(\ref{eq:gseven}) only has the trivial solution $A^{\dag}_l=0$ 
(the uniqueness of homogeneous and inhomogeneous problems involving linear 
differential operators on semi-infinite intervals can be proved by a 
transformation that takes the semi-infinite interval into a finite interval, 
followed by the utilization of theorems on uniqueness available for finite 
intervals. We provide a proof in Appendix A for the CGLE that is studied in 
detail in later sections. Moreover, such a transformation may also be
applied to operators with an asymptotic structure similar to that of 
the CGLE). This leads us to conclude that the modification of $A^{\dag}$ 
in a finite system is appropriately represented by Eq.~(\ref{eq:geight}).

For arbitrary functions $u$ (not the field $U$ in Eq.~(\ref{eq:gone})) 
and $v$, and using integration by parts, we have,
\begin{eqnarray}
(\pounds u,v)&=&(u,\pounds^{\dag}v)+v(b)u_x(b)-v(a)u_x(a)\nonumber\\
&+&v_x(a)u(a)-v_x(b)u(b),
\label{eq:gten}
\end{eqnarray}
where we assume for simplicity that $\pounds$ is a reaction-diffusion
type operator comprised of second order differential terms only.
$x=a$ and $x=b$ are arbitrary boundary points. If needed, one may evaluate 
surface terms for more general operators using integration by parts.

For the localized structure $a=-l$ and $b=\infty$. 
We invoke Eq.~(\ref{eq:geight}) and substitute 
$v=A_l^{\dag}$, $u=\delta U_l$ (the subscript $l$ denotes that $\delta U$ is
now considered in a finite system) in Eq.~(\ref{eq:gten}), to obtain,
\begin{eqnarray}
(\pounds \delta U_l,A_l^{\dag})&=&(f,A_l^{\dag})=
(\delta U_l,\lambda_l A_l^{\dag})+A_{lx}^{\dag}(-l)\delta U_l(-l)
\nonumber\\&-&A_{lx}^{\dag}(\infty)\delta U_l(\infty).
\label{eq:geleven}
\end{eqnarray}
This is the sought after finite system extension of the solvability criteria 
Eq.~(\ref{eq:gsix}). Also, as $l \rightarrow \infty$, Eq.~(\ref{eq:geleven}) 
reduces to $(f,A^{\dag})=0$. Since $\pounds$ is obtained by linearizing about 
the localized structure $U_0(x)$, $\delta U_l(-l)$ is simply the 
difference $U(-l)-U_0(-l)$, where $U(-l)$ is the Dirichlet boundary value 
imposed on field $U$, the solution of Eq.~(\ref{eq:gone}). 

The extension Eq.~(\ref{eq:geleven}), tailored to incorporate non-homogeneous 
Dirichlet boundary conditions on the field $U$, is not unique. For instance, 
one may consider the effects of non-homogeneous Neumann boundary conditions on 
the field $U$ by requiring that $A_l^{\dag}$ obeys
\begin{eqnarray}
\pounds^{\dag} A_l^{\dag}=\lambda_l A_l^{\dag},
~~A_{lx}^{\dag}(-l)=0,~~A_{lx}^{\dag}(\infty)=0.
\label{eq:gtwe}
\end{eqnarray}
Here, the derivatives, rather than $A_l^{\dag}$ itself, assume zero values at 
the boundary. Furthermore, an
extension $A_l^{\dag}$ for a general set of homogeneous boundary conditions, 
with homogeneous Dirichlet and Neumann boundary conditions as special cases, 
may also be developed. Next, we apply the techniques and criteria developed 
so far to analyze non-equilibrium Ising-Bloch fronts, as the fronts interact 
with the system boundary.                                                           
\section{Boundary effects in a generic Ising-Bloch system}
Ising-Bloch fronts provide an interesting arena to apply the methods developed 
in the last section. Along with the usual Goldstone mode associated with
translational invariance, the slow manifold for Ising-Bloch fronts also 
includes a spatially localized slow mode responsible for the Ising-Bloch
bifurcation \cite{skyrabin,michaelis,bode}. Chirality preserving stationary 
Ising fronts \cite{Coullet}, bifurcate into a pair of chirality broken, 
counter-propagating Bloch fronts. The slow manifold for Ising-Bloch fronts 
comprised of the Goldstone and chirality breaking modes, manifests itself in 
the form of order parameter equations (OPE) \cite{meron4,skyrabin,bode} for 
the order parameters, front velocity and front position. The front velocity is 
a measure of the effects of the chirality breaking mode. The Goldstone mode 
captures front translations by infinitesimal changes in the front position, 
the other order parameter. We seek the coupling between these order parameters
induced by the boundary data and broken translational invariance.

A generic Ising front denoted by $U_0(x)$, gives the Goldstone mode $U_{0x}$. 
Close to the Ising-Bloch bifurcation threshold, propagating Bloch wall 
solutions are regarded as perturbations of the stationary Ising wall 
solution \cite{Coullet}. The front velocity $c$ controls the strength of these 
perturbations. Therefore, expanding the deviation $\delta U$ in powers
of $c$, we have,
\begin{eqnarray}
U_b&=&U_0 +\delta U\nonumber\\
&=& U_0 +c\delta U_1 +c^2 \delta U_2 +c^3 \delta U_3 + ..,
\label{eq:h1}
\end{eqnarray}
with the perturbed Bloch wall solution $U_b$.  

For convenience we transform into a frame of reference moving along with
the Bloch wall. This transformation amounts to $\partial_t (\delta U)
\rightarrow \partial_t (\delta U)-c(U_{0x}+\delta U_x)$.
Invoking Eq.~(\ref{eq:gfive}) and substituting into it the expansion of
$\delta U$, while at the same time disregarding the influence of any external
perturbation $p(U,x)$, we obtain,
\begin{eqnarray}
&&\pounds[c\delta U_1+ c^2 \delta U_2+c^3 \delta U_3]=\partial_t(c \delta U_1)
\nonumber\\
&-&c[U_{0x} +c\delta U_{1x} +c^2 \delta U_{2x}]-c^2 N_2 -c^3 N_3+\cdots
\label{eq:h2}
\end{eqnarray}
$N_2$ and $N_3$ represent the coefficients of second order and third order
velocity terms respectively.

Equating terms which are first order in velocity $c$ in Eq.~(\ref{eq:h2}), we 
obtain,
\begin{eqnarray}
\pounds \delta U_1+U_{0x}=0.
\label{eq:h3}
\end{eqnarray}
This means that $\pounds$ has a double zero eigenvalue at the Ising-Bloch 
bifurcation threshold \cite{skyrabin,bode}. Therefore, along with the zero 
Goldstone mode, we have another eigenvalue that passes through zero at the
bifurcation. The Goldstone mode $U_{0x}$ and the 
generalized eigenvector $\delta U_1$ obtained from Eq.~(\ref{eq:h3}), span 
the slow manifold. The chirality breaking mode is then constructed as a 
linear combination of these two modes \cite{skyrabin}.

Employing the projection criteria Eq.~(\ref{eq:gsix}) for an Ising-Bloch front
close to the bifurcation threshold, ie., the solvability of Eq.~(\ref{eq:h2}),
results in,
\begin{eqnarray}
(\delta U_1,A^\dag)\partial_t c&=&c(U_{0x},A^\dag)
+c^2 (\delta U_{1x}+N_2,A^\dag)\nonumber\\
 &+& c^3 (\delta U_{2x}+ N_3, A^\dag) +\cdots
\label{eq:h4}
\end{eqnarray}
This is the generic OPE for the velocity
of Ising-Bloch fronts close to the bifurcation threshold. The particular form
of the inner products in Eq.~(\ref{eq:h4}) is system specific. If one
assumes further symmetries in the system, for example
$U\rightarrow-U$, inner products that are coefficients of even powers of the
velocity in Eq.~(\ref{eq:h4}) vanish, resulting in the normal form of a
pitchfork bifurcation. The inner product $(U_{0x},A^\dag)$ in 
Eq.~(\ref{eq:h4}) controls the distance from the Ising-Bloch bifurcation 
threshold, where for consistency (Ising-Bloch bifurcation is a pitchfork) it 
is further required that $(U_{0x},A^\dag) \sim c^2$, $\partial_t c \sim c^3$ 
\cite{skyrabin,bode}. Hence, all the terms in Eq.~(\ref{eq:h4}) are of size
$c^3$. 

We invoke the extended solvability criteria Eq.~(\ref{eq:geleven}) to 
evaluate the effects of boundary data on the dynamics of Ising-Bloch fronts.
For generic Ising-Bloch fronts interacting with boundaries where Dirichlet
data is present, the extended solvability criteria assumes the form,
\begin{eqnarray} 
(\delta U_{1l},A_l^{\dag})\partial_t c &=&c(U_{0x},A_l^\dag)
+c^2 (\delta U_{1lx}+N_2,A_l^\dag)\nonumber\\
&+& c^3 (\delta U_{2lx}+ N_3, A_l^\dag)\nonumber\\ 
&+& \lambda_l(c\delta U_{1l}
+c^2\delta U_{2l}+c^3\delta U_{3l}+\cdots,A_l^{\dag})\nonumber\\
&+& A_{lx}^{\dag}(-l)\delta U_{l}(-l)-A_{lx}^{\dag}(\infty)\delta U_{l}(\infty).
\label{eq:h5}
\end{eqnarray}
In contrast to earlier works \cite{meron4,skyrabin,bode} focused on the 
effects of external perturbations, $p(U,x)$, on the slow manifold, the 
constituent modes of the slow manifold require appropriate modifications in 
order to capture the effects arising due to confinement by boundaries.  
While, the modification of the adjoint Goldstone
mode $A^\dag$ to $A_l^{\dag}$ is generic to any confined localized 
structure, or alternatively, a localized structure being considered in the 
vicinity of system boundaries, the modification 
of the generalized eigenvector $\delta U_1$ to $\delta U_{1l}$ is a unique 
characteristic of Ising-Bloch fronts.

Simplifications to the slow manifold Eq.~(\ref{eq:h5}) are made by the 
following observations. Consider the term, 
$f_0=\lambda_l(c\delta U_{1l}+c^2\delta U_{2l}
+c^3\delta U_{3l}+\cdots,A_l^{\dag})$, on the right hand side of 
Eq.~(\ref{eq:h5}). The inner product 
$f_1=\lambda_l(c\delta U_{1l},A_l^{\dag})$ has the largest contribution since 
it involves the first power of the velocity $c$. Now, as mentioned 
before, all terms should be of size $c^3$, a requirement imposed for the 
Ising-Bloch bifurcation to be a pitchfork. 
Therefore, $f_1 \sim \lambda_l c \sim c^3$, implying
$\lambda_l \sim c^2$. Moreover, the size of $\lambda_l$ is controlled by the 
distance of the Bloch fronts from the boundary. If the front is far away from 
the boundary, that is, if $\lambda_l \sim {\cal O}(c^3)$, then 
$f_1 \sim {\cal O}(c^4)$, and its contribution to Eq.~(\ref{eq:h5}) can be 
neglected. As the front moves towards the boundary, so that 
$\lambda_l \sim c^2$, then $f_1 \sim c^3$ contributes to Eq.~(\ref{eq:h5}),
and the ensuing front dynamics. If the front gets too close to the boundary,
ie., $\lambda_l \sim c$, then $f_1 \sim c^2$, and the scaling requiring that 
all the terms be of size $c^3$ breaks down. In other words, if 
$\lambda_l \sim c$, the effects of the boundary are too strong for them to be 
accurately considered as small perturbations on the dynamics of Ising-Bloch 
fronts. Consequently, the size of $\lambda_l$ serves as a measure of the 
strength of the boundary perturbation. In light of the present discussion, 
Eq.~(\ref{eq:h5}) simplifies to
\begin{eqnarray}
(\delta U_{1l},A_l^{\dag})\partial_t c &=&c(U_{0x},A_l^\dag)
+c^2 (\delta U_{1lx}+N_2,A_l^\dag)\nonumber\\
&+& c^3 (\delta U_{2lx}+ N_3, A_l^\dag)\nonumber\\
&+& \lambda_l(c\delta U_{1l},A_l^{\dag})\nonumber\\
&+& A_{lx}^{\dag}(-l)\delta U_{l}(-l).
\label{eq:h6}
\end{eqnarray}
The surface terms at infinity contribute zero, since by construction
$A_l^{\dag}(\infty)=0$. 

\section{boundary effects in the parametrically forced CGLE}
The CGLE reads,
\begin{eqnarray}
\partial_\tau F=(\gamma+i\nu)F-|F|^2 F + \mu F^* + \partial{^2}_{X}F+ \alpha.
\label{eq:PCGLE}
\end{eqnarray}
Equation.~(\ref{eq:PCGLE}) and its generalizations \cite{Coullet,skyrabin,
ephlick} have been thoroughly analyzed in the context of the Ising-Bloch
bifurcation. The field $F$ may be regarded as the amplitude of diffusively
coupled auto oscillators that oscillate above the Hopf bifurcation threshold
determined by the parameter $\gamma$. $\mu$ represents the strength of
parametric forcing at twice the natural frequency, and $\nu$ is the detuning.
The parameter $\alpha$, which models forcing at the natural frequency of the
system, breaks the $(F\rightarrow -F)$ symmetry. As a result, the pitchfork
normal form of the Ising-Bloch bifurcation for $\alpha=0$ unfolds into a
saddle node for a non-zero $\alpha$.

We briefly recount the results of \cite{skyrabin} concerning 
the dynamics of Ising-Bloch fronts in the parametrically forced CGLE valid 
for an infinite system. This lays down the framework for the subsequent 
consideration of finite system sizes and boundary effects.

For $\alpha=0$ and in the bistable regime determined by the constraints,
$|\nu|<\mu$, $\gamma>-\sqrt{\mu^2-\nu^2}$, Eq.~(\ref{eq:PCGLE}) possesses 
a stationary Ising wall solution 
$F_I=\sqrt{\kappa}\tanh(\sqrt{\kappa/2X})e^{i\phi}$. Here $\kappa=\gamma
+\sqrt{\mu^2-\nu^2}$ and $\phi$ is obtained by solving $\sin(2\phi)=\nu
/\mu$. Bloch wall solutions of Eq.~(\ref{eq:PCGLE}) are then obtained as a
perturbation to the Ising wall, 
\begin{equation}
F_b(x,t)=\sqrt{\kappa}[\tanh(x)+u(x,t)+iw(x,t)]e^{i\phi},
\label{perturb}
\end{equation}
where the space-time scaling $t=\kappa \tau/2$, $x=\sqrt{\kappa/2}X$ is 
introduced by the authors, resulting in,
\begin{eqnarray}
\pounds =
\left[
\begin{array}{c}
D_1~~~~~~-4\nu/\kappa\\
\\
0~~~~~D_2-3+4\gamma/\kappa
\end{array}\right]\nonumber,\\ \nonumber \\ \nonumber \\
 D_1=\partial^2_x+2-6\tanh^2(x),\nonumber \\ \nonumber \\
 D_2=\partial^2_x+1-2\tanh^2(x),\nonumber\\ \nonumber\\
\widetilde{N}=-2\tanh(x)
~\left[
\begin{array}{c}
3u^2+w^2\\ \\
2uw
\end{array}\right]
&-2&\left[
\begin{array}{c}
u^3+uw^2\\ \\
w^3+wu^2
\end{array}\right]\nonumber.
\label{eq:donga}
\end{eqnarray}
For clarity and continuation of the conventions used in the previous 
sections, we stress the following points.
Firstly, we recognize that $\delta U=\{u,w\}^T$. Secondly, $\delta U$ obeys
\begin{eqnarray}
\partial_t \delta U=\pounds \delta U+\widetilde{N},
\end{eqnarray}
which when compared with Eq.~(\ref{eq:gfive}), leads to the realization that
$\widetilde{N}=N^{\prime \prime}(U_0){(\delta U)}^2/2+{\cal O}[{(\delta U)}^3$.
Thirdly, $\pounds$ is obtained by linearizing about the solution $U_0(x)$. 
In the present case the stationary solution is the
Ising wall $F_I(x)=\sqrt{\kappa}\tanh(x)e^{i\phi}$, and $U_0(x)=\tanh(x)$, 
where the constant factor $\sqrt{\kappa}e^{i\phi}$ should be dropped if the
perturbation $\delta U=\{u,w\}^T$ is defined through Eq.~(\ref{perturb}).

For the specific case of the parametrically forced CGLE, one has 
\cite{skyrabin},
\begin{eqnarray}
\delta U_1&=&\left[
\begin{array}{c}
\frac{8}{3\pi}I_{11}(x)-I_{12}(x)\\ \\ 
\frac{8\gamma}{9\pi\nu}$sech(x)$
\end{array}\right],~~~\nonumber
U_{0x}=\left[
\begin{array}{c}
$sech$^{2}(x)\\ \\ 
0
\end{array}\right],\nonumber\\ \\
\text{and}&~&~
A^\dag=\left[
\begin{array}{c}
\frac{9(\mu_c-\mu)\mu_c}{\pi\gamma\nu}$sech$^{2}(x)\\ \\ 
$sech(x)$
\end{array}\right].\nonumber
\label{eq:thevecs}
\end{eqnarray}
Substituting these vectors into Equation.~(\ref{eq:h4}) gives \cite{skyrabin},
\begin{eqnarray}
\partial_t c &=& \frac{27(\mu_c-\mu)\mu_c}{4\gamma^2} c
-\left({\left[\frac{8\gamma}{9\pi\nu}\right]}^2+0.36 \right) c^3.
\label{eq:opecgl}
\end{eqnarray}
Eq.~(\ref{eq:opecgl}) possesses three stationary 
states, two counter-propagating Bloch walls and a stationary Ising wall. These 
steady states exchange stability via the Ising-Bloch bifurcation at the 
critical bifurcation parameter $3\mu_c=\sqrt{9\nu^2+\gamma^2}$. 
The components of the vectors $\delta U=c \delta U_1+c^2 \delta U_2+..$, $U_0$ 
and $A^\dag$, in an infinite system, exponentially decay to zero as one
moves away from the front both to the left and to the right. This
signifies that Ising and Bloch walls are localized structures that are not
influenced by boundary conditions imposed on either boundary sufficiently 
far away. Furthermore, no explicit dependence on $x$ in Eq.~(\ref{eq:opecgl}) 
indicates translational invariance, a residue of infinite system size.

We now calculate $A^\dag_l$ and the associated value
of $\lambda_l$. $A^\dag_l$ satisfies the boundary conditions 
$A^{\dag}_l(-l)=0$, $A^{\dag}_l(\infty)=0$ (homogeneous problem), since we 
wish to examine the influence of Dirichlet boundary conditions on $U$
(non-homogeneous problem). Close to the bifurcation threshold determined by 
the magnitude of $\mu_c-\mu$, the operator $\pounds^\dag$ has the form
\begin{eqnarray}
\pounds^\dag&=&\left[
\begin{array}{c}
D_1~~~~~~~0\\ \\
-\nu/\gamma~~~~~D_2
\end{array}\right]\nonumber
+
\frac{27\mu_c(\mu-\mu_c)}{4\gamma^2}\left[
\begin{array}{c}
0~~~~~~~0\\ \\
\nu/\gamma~~~~-1
\end{array}\right]\nonumber\\ \nonumber\\
&=& \pounds^\dag_1+(\mu-\mu_c)\pounds^\dag_2.
\label{eq:matsum}
\end{eqnarray}
The operator $\pounds^\dag_2$ is a perturbative correction to the
operator $\pounds^\dag_1$, since $\mu-\mu_c \sim c^2$. Hence, we first 
examine $\pounds^\dag_1$ the dominant term in $\pounds^\dag$.

The operators $D_1$ and $D_2$ populate the diagonals of $\pounds^\dag_1$, and 
possess zero eigenvectors given by $Z_1=\text{sech}^2(x)$ and 
$Z_2=\text{sech}(x)$ respectively, in an infinite system. These 
eigenvectors satisfy the constraint of being zero at positive and negative 
infinity. Imagine a traveling Bloch front sufficiently distant from the 
boundary, where Dirichlet boundary conditions are imposed. The front does not 
sense the boundary and the condition $D_1Z_1=D_2Z_2=0$ holds. This is because 
the solutions $Z_1$ and $Z_2$ exponentially approach zero on either side of 
the front. As the front closes in on the boundary,
such that it is barely able to sense it ($Z_1$ and $Z_2$ have small finite 
values at the boundary), the eigenvectors $Z_1~\text{and}~Z_2$ 
are modified to $Z_{1l}~\text{and}~Z_{2l}$ by constraining them to 
have zero values at the boundary. Meanwhile, in a 
semi-infinite or finite domain, the only solutions to $D_1Z_{1l}=D_2Z_{2l}=0$ 
which have a zero value at both boundaries are the trivial solutions
$Z_{1l}=Z_{2l}=0$ (uniqueness arguments). Hence, requiring that the 
solutions $Z_{1l}$($Z_{2l}$) are only slight modifications of $Z_1$($Z_2$) and 
are not trivial zero solutions demands that these solutions 
obey $D_1Z_{1l}=\lambda_{1l}Z_{1l}$ and $D_2Z_{2l}=\lambda_{2l}Z_{2l}$.

\begin{figure}
\begin{center}
\begin{tabular}{cc}
\resizebox{80mm}{!}{\includegraphics[width=7cm,height=6cm,angle=0]
{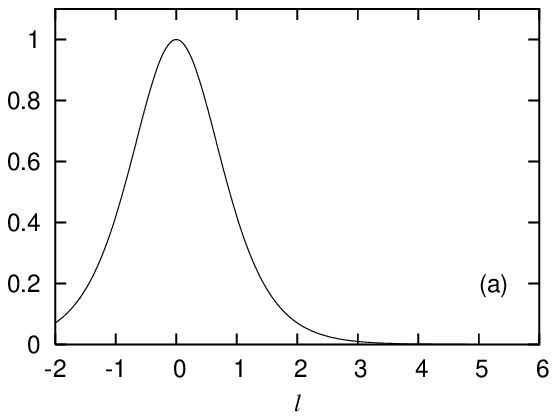}} \\
\resizebox{80mm}{!}{\includegraphics[width=7cm,height=6cm,angle=0]
{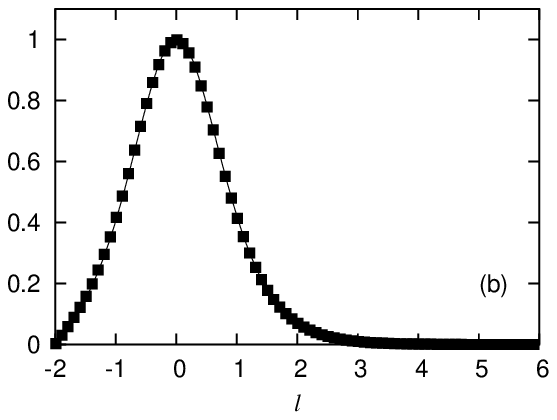}}
\end{tabular}
\caption{(a) Shows the plot of $Z_1$. The peak is at a distance of $l=2$
from the boundary. (b) The squares represent the numerically obtained
$Z_{1l}$. The analytical guess $Z_{1l}=\text{sech}^2(x)-\text{sech}^2(x+2l)$
is the solid line.}
\label{fig:eigvec}
\end{center}
\end{figure}

Figure.~\ref{fig:eigvec}(a) shows the plot of $Z_1$ in grey, where the left 
boundary is at a finite distance $l$ from the peak. $Z_1$ has a finite nonzero 
value at the boundary. We require that the modified eigenvector $Z_{1l}$ have a 
zero value at the boundary and not be all that different from $Z_1$ elsewhere.
We make the ansatz that this can be accomplished by subtracting from $Z_1$ 
its image to the left of the boundary. Therefore, we have, $Z_{1l}=
\text{sech}^2(x)-\text{sech}^2(x+2l)$. 
Figure.~\ref{fig:eigvec}(b) shows a good agreement between our guess and the
actual numerically evaluated $Z_{1l}$. This is so because in the asymptotic 
limit $\exp{2x}>>1$, $D_1=\partial_x^2-4$, and the image is approximately a 
zero eigenvector of this operator in the same limit.

Introducing images into a semi-infinite problem is by no means
a coincidence. Images are a common occurrence whenever boundary data is 
involved. For the extension $A^{\dag}_{l}$ (correspondingly 
$Z_{1l}$ and $Z_{2l}$) to assume a zero value at the boundary, the introduction
of the image becomes a natural necessity. Furthermore, we wish to stress that
the concept of images is quite general in its utility. Extensions of Goldstone
modes can be readily obtained for other systems, with linear operators having
similar properties of exponential decay asymptotics.
 
\begin{figure}
\begin{center}
\begin{tabular}{cc}
\resizebox{80mm}{!}{\includegraphics[width=7cm,height=6cm,angle=0]
{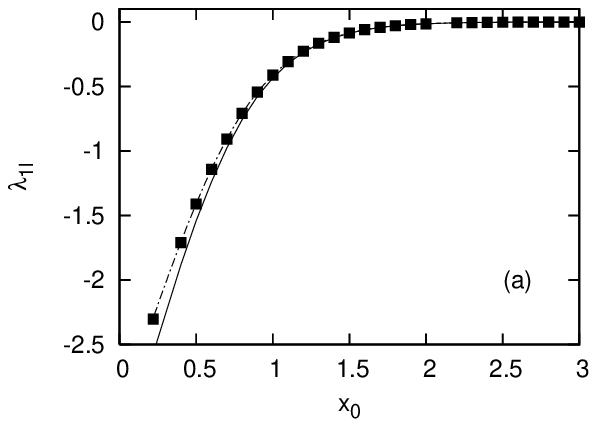}} \\
\resizebox{80mm}{!}{\includegraphics[width=7cm,height=6cm,angle=0]
{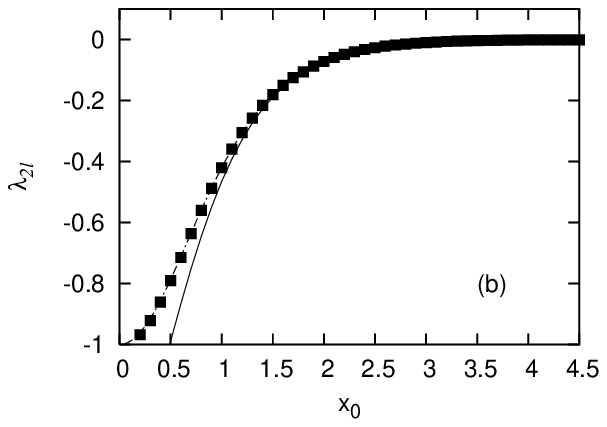}}
\end{tabular}
\caption{(a) Comparison of variational and numerical calculations of 
$\lambda_{1l}$
(b) Similar comparison of $\lambda_{2l}$ calculated using numerical and
variational techniques.}
\label{fig:eigenva}
\end{center}
\end{figure}

An upper bound, $\lambda_{1l}^\uparrow$, on the eigenvalue $\lambda_{1l}$, 
is easily obtained by a variational principle, given by, 
\begin{eqnarray}
|\lambda_{1l}|<|\lambda_{1l}^\uparrow|=(Z_{1l},D_1Z_{1l})/(Z_{1l},Z_{1l}).
\label{eq:varia}
\end{eqnarray}
A more refined variational guess of $Z_{1l}$ may be made 
by introducing an extra parameter $a_1$. Consequently, we 
have $Z_{1l}=\exp{(a_1x)}[\text{sech}^2(x)-\text{sech}^2(x+2l)]$. Manipulation 
of this parameter provides a better guess of the change in shape of the peak 
in the actual modified eigenvector $Z_{1l}$. Figure.~\ref{fig:eigenva}(a) 
compares the numerical and variationally calculated eigenvalues as a function 
of the distance $l$ of the front from the boundary. The dashed curve 
represents the numerically calculated eigenvalues of $D_1$. The thin curve
depicts the variationally calculated eigenvalues with $Z_{1l}=\text{sech}^2(x)-
\text{sech}^2(x+2l)$. The squares signify a better variational calculation
of the eigenvalues using $Z_{1l}=\exp{(a_1x)}[\text{sech}^2(x)-
\text{sech}^2(x+2l)]$. An improved guess of $Z_{2l}$, and eigenvalue 
$\lambda_{2l}$ for the operator $D_2$, similarly involves taking 
$Z_{l2}=\exp{(a_2x)}[\text{sech}(x)-\text{sech}(x+2l)]$. Depicted in 
Fig.~\ref{fig:eigenva}(b) are the eigenvalues $\lambda_{2l}$, numerically 
calculated (dashed curve), variationally calculated with respective guesses
$Z_{2l}=\text{sech}(x)-\text{sech}(x+2l)$ (thin line), and 
$Z_{2l}=\exp{(a_2x)}[\text{sech}(x)- \text{sech}(x+2l)]$ (squares).

The numerical calculation of the eigenvalues $\lambda_{1l}$
and $\lambda_{2l}$ involved using a standard QR algorithm on the matrix 
obtained by a finite difference approximation to the operators $D_1$ and 
$D_2$. The grid spacing was adjusted until we obtained convergence. The 
eigenvectors were calculated using inverse iterations, with the number of 
iterations optimized for convergence.

The first row in the matrix representation of the adjoint operator 
Eq.~(\ref{eq:matsum}) consists only of the operator $D_1$. Therefore, 
since $\pounds^\dag A^\dag_l=\lambda_l A^\dag_l$, we immediately 
obtain $\lambda_l=\lambda_{1l}$. We recall 
that in the limit of infinite front distance from the boundary 
$l\rightarrow\infty$, we have $A^\dag_l\rightarrow A^\dag$. Combining this 
asymptotic limit constraint with the requirement that the sought after 
eigenvector has zero values at both boundaries, we obtain,
\begin{eqnarray}
A^\dag_l=\left[
\begin{array}{c}
\frac{(\mu_c-\mu)\mu_c}{\pi\gamma\nu}
Z_{1l}  \\ \\ 
Z_{2l}
\end{array}\right].\nonumber\\
\label{eq:advec}
\end{eqnarray}
A more rigorous derivation involving a step by step consideration of the 
operators $L^\dag_1$ and $L^\dag_2$ in a perturbative scheme also yields 
Eq.~(\ref{eq:advec}).

We now focus on incorporating the effects of the Dirichlet boundary values
$X_b$ and $Y_b$, the values of the real and imaginary components of the
field $F$ in Eq.~(\ref{eq:PCGLE}), into the dynamics of fronts close to the
boundary. Bloch walls are perturbed Ising walls, with the
perturbation $\delta U_l$. The boundary value of this perturbation 
$\delta U_l(-l)$ is obtained by fixing $F(-l)=X_b+iY_b$ and subtracting from 
it the value that the Ising wall 
assumes $F_I(-l)=\sqrt{\kappa}\tanh(-l)e^{i\phi}$. Recalling
Eq.~(\ref{perturb}), and $\delta U=\{u,w\}^T$, we obtain, 
\begin{eqnarray}
\delta U_l(-l)=\left[
\begin{array}{c}
( X_b $cos$(\phi)+Y_b $sin$(\phi))/\sqrt{\kappa}+$tanh$(l) \\ \\ 
( Y_b $cos$(\phi)-X_b $sin$(\phi))/\sqrt{\kappa} 
\end{array}\right].\nonumber \\
\label{eq:bdev}
\end{eqnarray}

\section{OPE}

To extract a reduced description of the influence of Dirichlet boundary 
conditions on the motion of Ising-Bloch fronts, we invoke Eq.~(\ref{eq:h6}), 
and substitute into it the explicit forms of $A^\dag_l$ and $\lambda_l$ 
derived in the previous section. Consider the term 
$f_1=\lambda_l(c\delta U_{1l},A_l^{\dag})$ on the right hand 
side (RHS) of Eq.~(\ref{eq:h6}). For the CGLE, as seen in 
Eq.~(\ref{eq:advec}), the first component of $A_l^{\dag}$, denoted by,
$A_{l1}^{\dag }$, is smaller by a factor of $c^2$ than the second component
$A_{l2}^{\dag }$. This is so because $\mu_c -\mu \sim c^2$. Hence, while 
evaluating $f_1$, we need only consider the inner product of the second 
component of the generalized eigenvector, $\delta U_{1l}$, denoted by 
$\delta U_{1l2}$, and $A_{l2}^{\dag }$. The generalized eigenvector 
$\delta U_1$ is known Eq.~(\ref{eq:thevecs}), and its finite system 
modification $\delta U_{1l}$  
needs to be evaluated (only the second component $\delta U_{1l2}$) to 
evaluate the inner product in $f_1$. 

To evaluate $\delta U_{1l2}$ we recall that $Z_2=\text{sech}(x)$, 
with $D_2Z_{2}=0$. The second component of
$\delta U_1$, is given by $\delta U_{12}=[8\gamma/9\pi\nu]\text{sech}(x)$.
Hence, $D_2 \delta U_{12}=0$.
In a confined system with the left boundary at $x=-l$, $Z_2$ is modified to
$Z_{2l}=\text{sech}(x)-\text{sech}(x+2l)$, requiring that the
homogeneous boundary condition, $Z_{2l}(-l)=0$, holds good. In the confined
system $\delta U_{12}$ is modified to $\delta U_{1l2}$. However, to obtain
$\delta U_{1l2}$, the requirement that it obeys the inhomogeneous boundary
condition $c\delta U_{1l2}(-l)=\delta U_{l2}(-l)$, since $\delta U_l=
c\delta U_1 +{\cal O}(c^2)$, needs to be imposed. Therefore we construct
$\delta U_{1l2}(x)=c\delta U_{12}-\beta \text{sech}(x+2l)$, followed by
imposing the inhomogeneous boundary condition
$c\delta U_{1l2}(-l)=\delta U_{l2}(-l)$, to evaluate $\beta$. After doing so,
we have,
\begin{eqnarray}
c \delta U_{1l2}=\frac{c8\gamma}{9 \pi \nu} Z_{2l}
-\frac{\delta U_{l2}(-l)}{\text{sech}(l)}\text{sech} (x+2l).
\label{eq:modi}
\end{eqnarray}

We, finally have the ingredients to calculate all the inner products in
Eq.~(\ref{eq:h6}). The bulk of the boundary
influence, we contend, is captured by the interplay of the terms,
$c(U_{0x},A_l^\dag)$, $\lambda_l(c\delta U_{1l},A_l^{\dag})$, and
the surface term $A_{lx}^{\dag}(-l)\delta U_{l}(-l)$ in Eq.~(\ref{eq:h6}). 
Therefore, although, strictly speaking, the inner products containing higher 
order terms $c^2 (\delta U_{1lx}+N_2,A_l^\dag)$, and 
$c^3 (\delta U_{2lx}+ N_3, A_l^\dag)$, in Eq.~(\ref{eq:h6}), should be
evaluated in the finite domain $[-l,\infty]$, we approximate them by taking
the inner product in the infinite interval $[-\infty,\infty]$.

Performing all the inner products in Eq.~(\ref{eq:h6}) and rearranging the
terms, we obtain
\begin{eqnarray}
\partial_t c &=& \frac{27(\mu_c-\mu)\mu_c}{4\gamma^2} c+\lambda_l c
-\left({\left[\frac{8\gamma}{9\pi\nu}\right]}^2+p\right) c^3\nonumber\\
&-&\left[\frac{9\pi\nu}{16\gamma}\right]\text{tanh}(l)
\text{sech}(l) \delta U_{l2}(-l)\nonumber\\
&+&\left[\frac{81(\mu_c-\mu)\mu_c}{4\gamma^2}\right]
\text{tanh}(l) \delta U_{l1}(-l)\nonumber\\
&-&\left[\lambda_l\frac{9\pi\nu}{16\gamma}\right]2l\text{cosech}(2l).
\label{eq:dopecgl}
\end{eqnarray}
In deriving Eq.~(\ref{eq:dopecgl}) 
we have used $Z_{1l}=\text{sech}^2(x)-\text{sech}^2(x+2l)$ and
$Z_{2l}=\text{sech}(x)-\text{sech}(x+2l)$, where $\lambda_l=\lambda_{1l}$ is 
given by Eq.~(\ref{eq:varia}), and $p=0.36$ Eq.~(\ref{eq:opecgl}). 
Equation.~(\ref{eq:dopecgl}) along with 
$\partial_t l=-c$ represents the coupling of the two degrees of freedom, 
front velocity $c$ and position $l$, by the influence of Dirichlet boundary 
conditions imposed at the boundary. As required, in the limit of infinite front distance from the boundary Eq.~(\ref{eq:dopecgl}) reduces to 
Eq.~(\ref{eq:opecgl}). 

We now examine the consequences of the coupling of the front velocity and
position close to the boundary. Firstly, we report the findings of our 
numerical simulations of Eq.~(\ref{eq:PCGLE}), which is a system with infinite
degrees of freedom. Secondly, we corroborate these findings by solving the 
reduced, two degree of freedom OPE we have derived.
  
We performed numerical simulations of Eq.~(\ref{eq:PCGLE}), where Bloch fronts
were created at infinity (far from the boundaries) and launched towards a 
boundary. The velocity of these Bloch fronts was chosen to be one of the
steady states of Eq.~(\ref{eq:opecgl}) resulting in uniform front translation
with this velocity until the fronts closed in on the boundary. Near the 
boundary, contingent upon 
the Dirichlet boundary value imposed, the incoming Bloch fronts were either 
trapped or bounced back. Bloch fronts that bounce evolve into the 
counter-propagating Bloch front near the boundary and move away. Trapped Bloch
fronts, as opposed to bouncing Bloch fronts, evolve into non-trivial steady 
state solutions (See Ref.\cite{yadav}) of the CGLE Eq.~(\ref{eq:PCGLE}).

\begin{figure}
\includegraphics[width=8cm,height=8cm,angle=0]{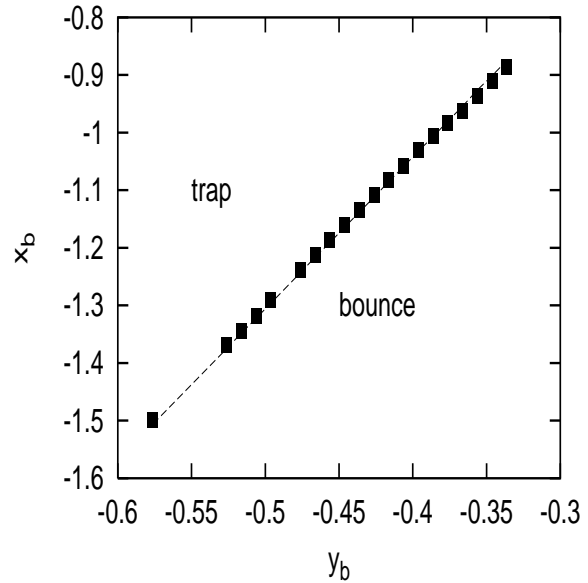}
\caption{The transition curve for the full model Eq.~(\ref{eq:PCGLE}) plotted
using squares, the same curve obtained from the reduced OPE  
Eq.~(\ref{eq:dopecgl}), plotted as a dashed line. Here, 
$\nu=0.3$, $\gamma=1.0$, $\mu=0.448$.} 
\label{fig:trans}
\end{figure}

We summarize our numerical observations of Bloch front behavior as a function 
of the boundary conditions $X_b$ and $Y_b$ in Figure.~\ref{fig:trans} . 
This phase diagram in the plane of 
boundary values reveals a curve separating regions of bouncing and trapped
fronts represented by diamonds. We compare these results with the 
transition curve predicted by the reduced model Eq.~(\ref{eq:dopecgl}), 
plotted as the dashed curve in Figure.~\ref{fig:trans}. 
The plots show a good agreement (within $0.5\%$)
between the two transition curves. This is a striking result considering the
fact that in calculating $A^\dag_l$ and $\lambda_l$ we have employed 
approximate vectors $Z_{1l}$ and $Z_{2l}$.

Bouncing fronts gradually slow down as they near the boundary, attain zero
velocity at a certain critical distance from it, and finally move away as 
the sign of the velocity flips. 
As we change the boundary values and get closer to
the transition curve, bouncing fronts attain zero velocity at a much smaller
critical distance from the boundary, until eventually right at the transition
curve they reach the point of closest approach to the boundary. As we 
cross the transition curve and move into the trapping region, approaching 
fronts no longer attain zero velocity close to the boundary, their velocity 
never flips sign, and hence they never bounce. The distance from the boundary
of the point of closest approach depends on where exactly on the phase 
diagram the transition curve is crossed.

The agreement between the transition curves obtained from the full model
Eq.~(\ref{eq:PCGLE}) and the reduced model Eq.~(\ref{eq:dopecgl}) is better
when the point of closest approach is further away from the boundary. This is
because, as detailed earlier, the vectors $Z_{1l}$ and $Z_{2l}$ are better
approximations to the actual solutions of $D_1Z_{1l}=\lambda_1 Z_{1l}$ and 
$D_2Z_{2l}=\lambda_2 Z_{2l}$, further away from the boundary. Consequently, 
a better guess of these vectors, valid close to the boundary, should improve 
the agreement between the transition curves, even if, the point of closest 
approach is closer to the boundary. However, the approximate vectors we 
use are sufficient for the purpose of establishing the usefulness of our 
general method that accounts for the broken translational invariance in a 
spatially finite system through the extension of solvability conditions. Our
 method incorporates into it the eigenvalue $\lambda_l$, the most direct 
measure of broken translational invariance, which can be obtained accurately 
via a variational principle using relatively crude guesses for the
eigenvectors.                       

\begin{figure}
\begin{center}
\begin{tabular}{cc}
\resizebox{80mm}{!}{\includegraphics[width=7cm,height=6cm,angle=0]
{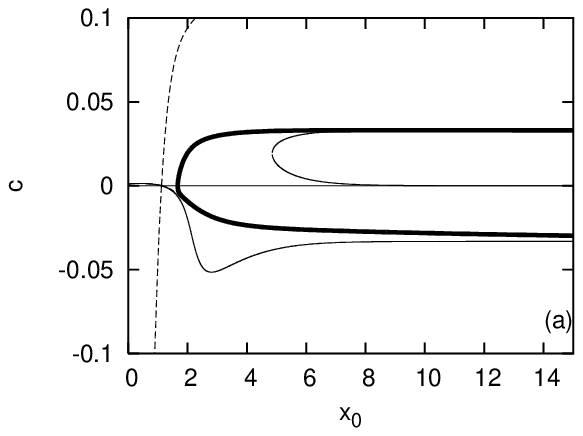}} \\
\resizebox{80mm}{!}{\includegraphics[width=7cm,height=6cm,angle=0]
{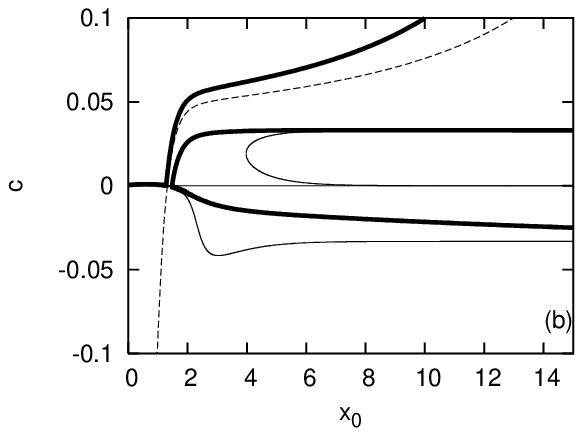}}
\end{tabular}
\caption{(a) The plot deep inside the bouncing region, the nullclines are
thin black curves, the thick curves correspond to the trajectories in the
phase plane, and the invariant manifolds are plotted as dashed lines. Here, 
$\nu=0.3$, $\gamma=1.0$, $\mu=0.448$, $X_b=-1.116$, and $Y_b=-0.4262$. 
(b) Plot still in the bouncing region, but close to the transition curve. The 
same plotting scheme and parameters used, with boundary values 
$X_b=-1.112$, $Y_b=-0.4262$.}
\label{fig:bounce1}
\end{center}
\end{figure}

We now, by examining Eq.~(\ref{eq:dopecgl}) in more detail, extract the 
mechanism behind the transition from bouncing to trapped fronts as Dirichlet 
boundary conditions are changed. Figure.~\ref{fig:bounce1}(a) 
shows the nullclines, invariant
manifold, and trajectories of Eq.~(\ref{eq:dopecgl}) inside the bouncing
region of the phase diagram. A saddle, present at the point of intersection of
the nullclines, controls the flows in this bouncing regime.  Far away from the
boundary, situated at $x=0$ in the plot, the nullclines are three parallel
straight lines that represent two counter-propagating Bloch wall steady state
solutions, and a stationary Ising wall solution of Eq.~(\ref{eq:opecgl}).
The bouncing involves the Bloch front initially flowing towards the saddle.
Thereupon, influenced by the unstable manifold, the front flows away.

Figure.~\ref{fig:bounce1}(b)  
still depicts flows inside the bouncing region, but much closer to
the transition curve. In this regime bouncing and trapped fronts can coexist.
The invariant manifolds demarcate two basins, one of attraction towards the
boundary, and the other of repulsion away from it. Inside the repulsion basin
all incoming Bloch fronts bounce with the same mechanism as in 
Fig.~\ref{fig:bounce1}(a).
All the flows in the attraction basin are directed towards the system boundary,
with no possibility of a bounce. Figure.~\ref{fig:bounce1}(b) 
shows both bouncing and trapped
Bloch front trajectories in their respective basins. We reported on the
the coexistence region in our numerical study of Eq.~(\ref{eq:PCGLE}) in
Ref.\cite{yadav}. Here, we have provided an analytical explanation of this 
phenomena.

\begin{figure}
\begin{center}
\begin{tabular}{cc}
\resizebox{80mm}{!}{\includegraphics[width=7cm,height=6cm,angle=0]
{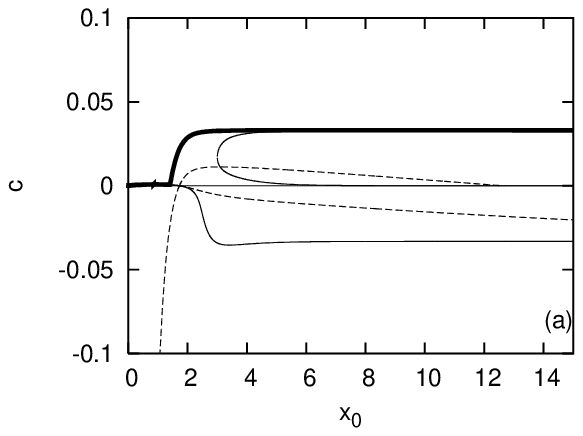}} \\
\resizebox{80mm}{!}{\includegraphics[width=7cm,height=6cm,angle=0]
{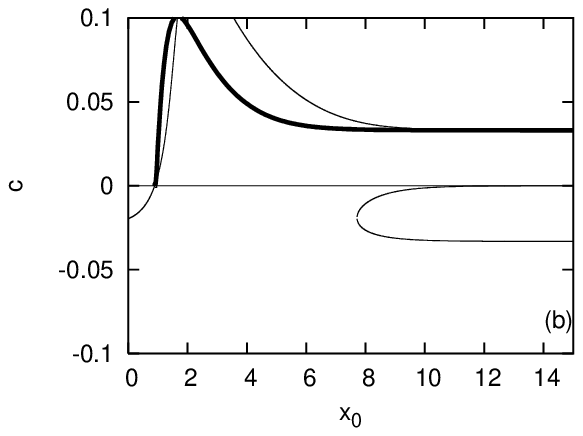}}
\end{tabular}
\caption{(a) Plot in the trapping
region close to the transition curve. The same plotting scheme and
parameters used, with boundary values $X_b=-1.11$, $Y_b=-0.4262$.
(b) The plot deep inside the trapping region, the nullclines are
thin black curves, the trajectory is the thick curve. Here,
$\nu=0.3$, $\gamma=1.0$, $\mu=0.448$, $X_b=-1.09$, and $Y_b=-0.4262$.}
\label{fig:trap1}
\end{center}
\end{figure}

The flows in the trapping region close to the transition curve are shown in
Figure.~\ref{fig:trap1}(a) . 
Trapped Bloch fronts, created at infinity and on the upper branch 
of the nullcline (corresponding to one of the steady states of
Eq.~(\ref{eq:opecgl})), lie inside the basin of attraction towards the 
boundary. Consequently, the transition from bouncing to trapped fronts is 
marked by the initial front velocity and position moving from the basin of 
repulsion (Fig.~\ref{fig:bounce1}(b)) to the 
basin of attraction (Figure.~\ref{fig:trap1}(a)) as the boundary values 
are varied. Deep inside the trapping region the saddle no longer exists, 
and we have a sink instead (Fig.~\ref{fig:trap1}(b)). All incoming Bloch 
front trajectories end up at this sink.

Summarizing, the nonuniform motion of Bloch fronts close to the boundary
is governed by the fixed point of Eq.~(\ref{eq:dopecgl}), giving rise to
bouncing, trapping, and coexistence of the two. Well inside the
bouncing region this fixed point is a saddle. Deep into the trapping region
the fixed point changes into a sink.

\section{Conclusion}
We have developed a general method of analyzing the influence of broken
translational invariance due to finite size and boundary effects on the 
dynamics of localized
solutions of generic non-linear spatially extended systems. We apply our
method to the special case of a bistable reaction-diffusion system, where
the localized solutions are fronts Eq.~(\ref{eq:dopecgl}). The implementation
of this method involves the extension of the infinite system size limit 
solvability conditions, used to extract a reduced description of the infinite
dimensional system, into solvability conditions that account for finite system
size and boundary effects. The extended solvability criteria works by naturally 
incorporating into it the concept of images. As a result, the method affords
a direct grasp of the broken translational invariance in a confined system
through the calculation of relevant eigenvalues.

In the special case of Dirichlet boundary conditions imposed on the CGLE,
we were able to provide mechanisms for Bloch front trapping, bouncing and
coexistence of the two at the boundary. This nonuniform front motion
is a result of the coupling of the two degrees of freedom, front velocity and
position, by the influence of boundary conditions. We have explicitly derived
this coupling by using our method of solvability condition extension. The
role of other types of boundary conditions, either Neumann or mixed can be
explored in a similar fashion by constructing a suitable extension of the
modified Goldstone mode. For example, exploring Neumann boundary conditions
requires the extension to always have zero derivatives at the boundary. 
This can be accomplished in the CGLE or other systems by adding, rather than 
subtracting, the image.

Finally, we comment on the generality of solvability condition extension. 
In any system, whenever it is possible to derive reduced dynamical equations 
through projections on the Goldstone mode, our method can be applied to 
obtain the finite size and boundary effects in terms of the modifications of 
these reduced dynamical equations. 

\begin{acknowledgements}
This work was supported in part by NSF Grant No. DMR-9710608 and by a Faculty
Research Grant from the Louisiana State University office of Sponsered
Research.
\end{acknowledgements}

\appendix

\section{}

For the CGLE, consider the operator $D_1$ Eq.~(\ref{eq:donga}) in a 
semi-infinite interval $[-l,\infty]$. Using the transformation 
$t =1-e^{-(l+x)}$, the problem
\begin{eqnarray}
D_1 Y&=&[\partial^2_x+2-6\tanh^2(x)] Y=0,\nonumber \\
Y(-l)&=&0~;~Y(\infty)=0,
\label{eq:a2}
\end{eqnarray}
is transformed to
\begin{eqnarray}
[\partial_t^2&-&\frac{\partial_t}{(1-t)}+
\frac{2+6\tanh^2(l+\ln(1-t))}{{(1-t)}^2}]Y=0 ,\nonumber\\
Y(0)&=&0~;~Y(1)=0.
\label{eq:a3}
\end{eqnarray}
Equation.~(\ref{eq:a3}) has a regular singular point at $t=1$, and thus
has a unique solution. Similar considerations apply to the operator $D_2$.
Therefore, homogeneous or inhomogeneous problems involving the operator
$\pounds$, which is comprised of the operators $D_1$, and $D_2$, should have 
unique solutions in a semi-infinite domain. For operators that
possess exponential decay asymptotics (true for a wide variety of models of
physically occurring localized structures), a transformation of the type used
here, can always be found in order to prove the uniqueness.

\end{document}